# Spin Neurons: A Possible Path to Energy-Efficient Neuromorphic Computers

Mrigank Sharad, D. Fan, and Kaushik Roy

School of Electrical and Computer Engineering, Purdue University, West Lafayette, Indiana 47907, USA

Recent years have witnessed growing interest in the field of brain-inspired computing based on neural-network architectures. In order to translate the related algorithmic models into powerful, yet energy-efficient cognitive-computing hardware, computing-devices beyond CMOS may need to be explored. The suitability of such devices to this field of computing would strongly depend upon how closely their physical characteristics match with the essential computing primitives employed in such models. In this work we discuss the rationale of applying emerging spin-torque devices for bio-inspired computing. Recent spin-torque experiments have shown the path to low-current, low-voltage and high-speed magnetization switching in nano-scale magnetic devices. Such magneto-metallic, current-mode spin-torque switches can mimic the analog summing and 'thresholding' operation of an artificial neuron with high energy-efficiency. Comparison with CMOS-based analog circuit-model of neuron shows that spin neurons can achieve more than two orders of magnitude lower energy and beyond three orders of magnitude reduction in energy-delay product. The application of spin neurons can therefore be an attractive option for neuromorphic computers of future.

## I. Introduction

Several neural-network based computing models have been explored in recent years for realizing hardware that can perform brain-like cognitive-computing [1]. The fundamental computing units of such-systems can be identified as the 'neurons' that connect to each other and to external stimuli through adaptable or programmable connections called 'synapses' [1]. Large number of neurons can be connected in various different network-topologies to realize different neural-network architectures. For instance, cellular-neural-networks employ near-neighbor connectivity [2], whereas, feed-forward networks employ all-to-all connections between neurons in consecutive network-stages (fig. 1a) [3]. Several other network-paradigms like Convolution Neural Networks (CNN) [4], and Hierarchical Temporal Memory (HTM) [5], may employ pyramidal interconnections in which a larger number of neurons in a lower-level of network connect to fewer neurons at the next higher-level. The more recent network models possess higher learning capacities and are capable of performing more complex cognitive-computing tasks [5].

Irrespective of the network-topology, the energy-efficiency, the performance and the integration density of neuromorphic hardware would be governed by the design of the fundamental computing units, i.e., the neurons. The basic operation of a step-transfer function neuron can be expressed as a 'sign' or 'threshold' operation given by eq.1 [1].

$Y = \text{sign} (\Sigma W_i I_i + b_i)$  (1)

Where, $I_i$ denote the $i_{th}$ input to the neuron, $W_i$ the corresponding synapse-weight and $b_i$ the neuron-bias. The input-weights (that can be positive or negative) can be realized using compact programmable, non-volatile resistive elements, namely memristors [6]. Several different device-techniques, including spintronic-memristors [7], have been proposed and demonstrated in literature [8]. Application of input voltages to such resistive input-weights results in analog-currents that are summed and compared with a threshold (which can be zero) by the neuron. Conventionally, CMOS operation amplifiers have been used in literature for implementing the analog summation and thresholding operation of neurons [3, 7]. However, such schemes may not lead to scalable and energy-efficient designs. We proposed the application of ultra-low-voltage, current-mode spin-torque switches as neurons in our recent work [9-12].. In this work we present the basic rationale of using nano-scale spin-torque switches as 'neurons' for the design of highly energy-efficient neural-networks [10]. We discuss the specific device-characteristics of such spin-torque switches that lend themselves to an efficient mapping of the neuron-equation (eq.1). We show how the terminal characteristics of spin-neurons can provide more than three orders of magnitude reduction in energy-delay product as compared to the conventional CMOS circuits.

## II. Conventional Neuron Circuit

Fig. 1b depicts an ideal circuit-model for a neuron with step transfer-function given by eq. 1. The synapse-weights are implemented using programmable conductance elements $G_i$ (which can potentially have negative values). Input voltages $V_i$ applied to the synapses result in a current $\Sigma G_i V_i$, which can be either positive or negative, depending upon the set of inputs and the weights. The neuron-output, acting as a current-dependent binary voltage-source, assumes a high (+1) or a low (-1) value, depending upon the sign of the total current. It is important to note the essential input characteristics provided by the idea neuron model. The input port provides a fixed potential (in this case, ground potential) and offers small input impedance (ideally zero).

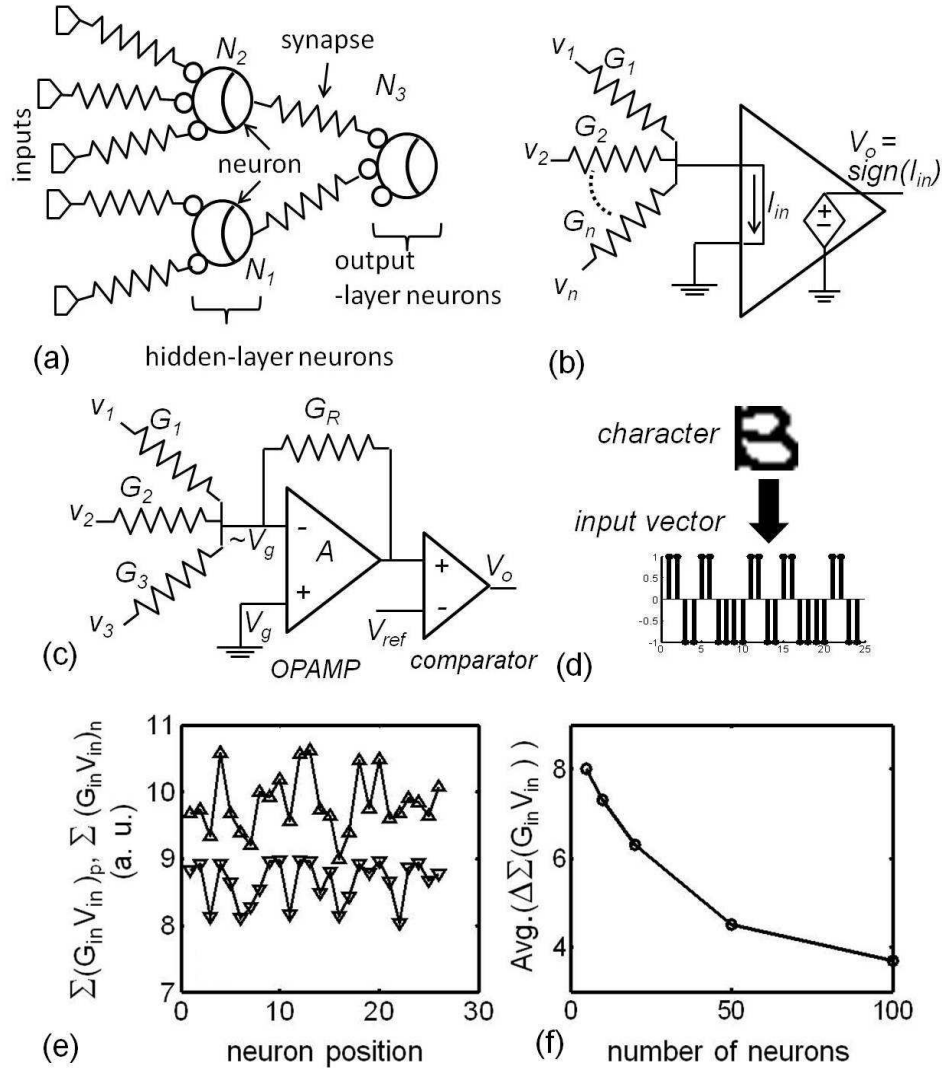

Fig.1 (a) A feed-forward Neural Network constituting of multiple neurons, (b) an ideal circuit model for step-transfer function neuron, (c) an analog CMOS realization neuron., (d) input vector generation from character images using method described in [9], (e) $|\Sigma(G_i V_i)_p|$ and $|\Sigma(G_i V_i)_p|$ values for 26 output neurons for character-recognition operation, (f) $\Delta VG_i$ vs. number of neurons.

This essentially implies that there is negligible change in the voltage potential at the input port. Note that any significant deviation in the input potential from a desired value would result in a net current of $\Sigma G_i (V_i - V_{in})$, where $V_{in}$ is the non-zero input potential. This would cause erroneous network outputs when $V_{in}$ varies randomly for different neurons.

A practical CMOS circuit design to implement the ideal neuron model presented in fig. 1b is given in fig. 1c. An operation amplifier (OPAMP) is used at the first stage of the circuit, which, for a sufficient amplification-gain, forces its two inputs to remain close to each other. Thus, by applying a fixed voltage on one of the two inputs (ground-potential $V_g$), the other input, (which is used as the neuron-input terminal) is also clamped to the same potential. Assuming $V_g = 0$, the output voltage of the OPAMP can be visualized as $Vo = (1/G_R)\Sigma G_i (V_i)$, which can be positive or negative. The result is compared with zero using a comparator. For an appropriate choice of $G_R$ the output voltage swing can be made sufficiently large so that a simple inverter can be used as a comparator in the second stage.

This example shows that the conventional circuit model of neuron employs an OPAMP for providing a low-impedance (fixed-voltage) input-node for linear summation of input-currents, and for transimpedance conversion of the current-mode summation, to yield the neuron output. Thus the energy-efficiency and the performance of such a neuron model would be limited by the characteristics of the OPAMP, which is a power and area consuming circuit.

The summation term in eq. 1 can be divided into its positive ($\Sigma(G_i V_i)_p$) and negative ($\Sigma(G_i V_i)_n$) constituents. The result of the sign operation is determined by the difference between these two terms ($|\Sigma(G_i V_i)_p| - |\Sigma(G_i V_i)_n|$), which is essentially $\Sigma(G_i V_i)$. As an example, we obtained the network parameters for a 2-layer feed-forward neural network for character recognition using the method described in [9]. The output layer of the network has 26 neurons, each corresponding to one of the 26-alphabetic characters. Fig. 1e shows the plot for $|\Sigma(G_i V_i)_p|$ and $|\Sigma(G_i V_i)_n|$ for the 26-neurons for the case when the input character belongs to the particular nodes. Results show that $\Sigma(G_i V_i)$ can be less than 10% of the total positive and negative current ( denoted by $|\Sigma(G_i V_i)_p| + |\Sigma(G_i V_i)_n|$) flowing through the synapses. Thus, the resolution required for the neuron for correct operation can be defined as the ratio given in eq. 2

$$\Delta VG_i = (|\Sigma(G_i V_i)_p| - |\Sigma(G_i V_i)_n|) / |\Sigma(G_i V_i)_p| + |\Sigma(G_i V_i)_n| \times 100 \qquad (2)$$

Fig. 1f shows that $\Delta VG_i$ for a neuron reduces with increasing number of inputs. For neurons with larger than ~25 inputs, this value can be lower than ~5%. This translates to stringent constraints upon the variations in the input voltage of the neuron. As mentioned above, any random variation in the bias voltage of the input port would result in deviation from the ideal neuron equation, resulting in computing errors.

Results show that after considering 10% $\sigma$ variations in the input weights, we are left with less than 3% tolerance for the variation in the input node-voltage. For OPAMP supply as well as the binary-input level of $\pm$ 0.5 V in 45nm technology, this would translate to ~30mV of tolerance. Notably, the random offsets in an OPAMP can be few tens of millivolts [13]. The sizing and gain of the OPAMP must be large enough to meet the offset requirements. With the aforementioned constraints, we obtained the power-consumption, delay, energy (per-operation) and energy-delay product for a 25-input CMOS neuron-circuit shown in fig. 1c, for different supply voltages (rail to rail). The results are given in fig. 2 a-d. At the optimal point, power-consumption and the bandwidth (delay-1) were found to be around ~70μW and ~100MHz respectively. This provided an optimal energy-dissipation of ~0.7pJ per-neuron per-cycle. The energy-delay-product can be obtained as ~$3.5e^{-21}$ J-s. The maximum current per-synapse used for this case was ~3μA.

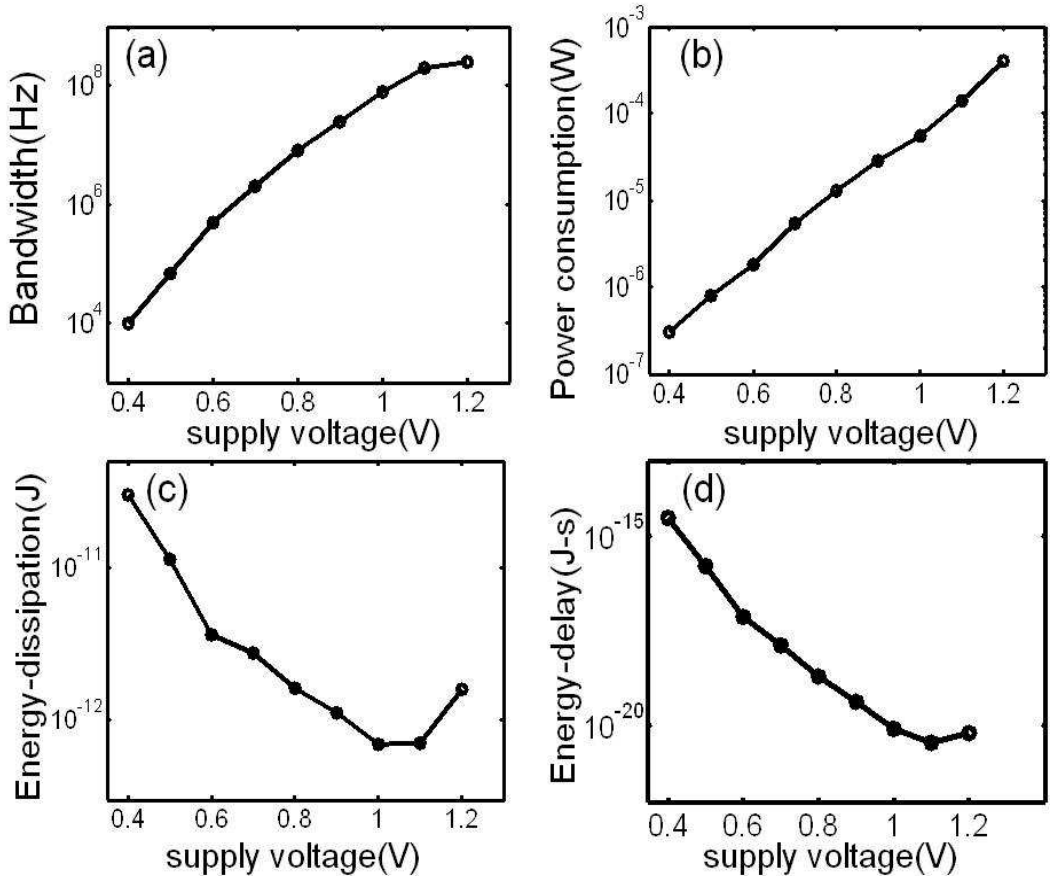

Fig. 2 (a) Bandwidth of CMOS neuron circuit vs. supply voltage, (b) power consumption vs. supply voltage, (c) energy-dissipation per computing operation of CMOS neuron vs. supply voltage, (d) energy-delay product vs. supply voltage.

Notably variability-related design constraints may become increasingly more stringent at lower technology nodes for conventional analog circuits, leading to heightened design challenges.

We next present the design and analysis of spin-torque based neuron and discuss its energy-benefits over CMOS model discussed above.

### III. Spin Torque Neuron

In our recent work, we proposed the application of spin-torque neurons for designing ultra-low power neural networks. Application of device structures based on lateral spin valves [9, 11], as well as domain-wall magnets (DWM) [10, 12] were proposed.

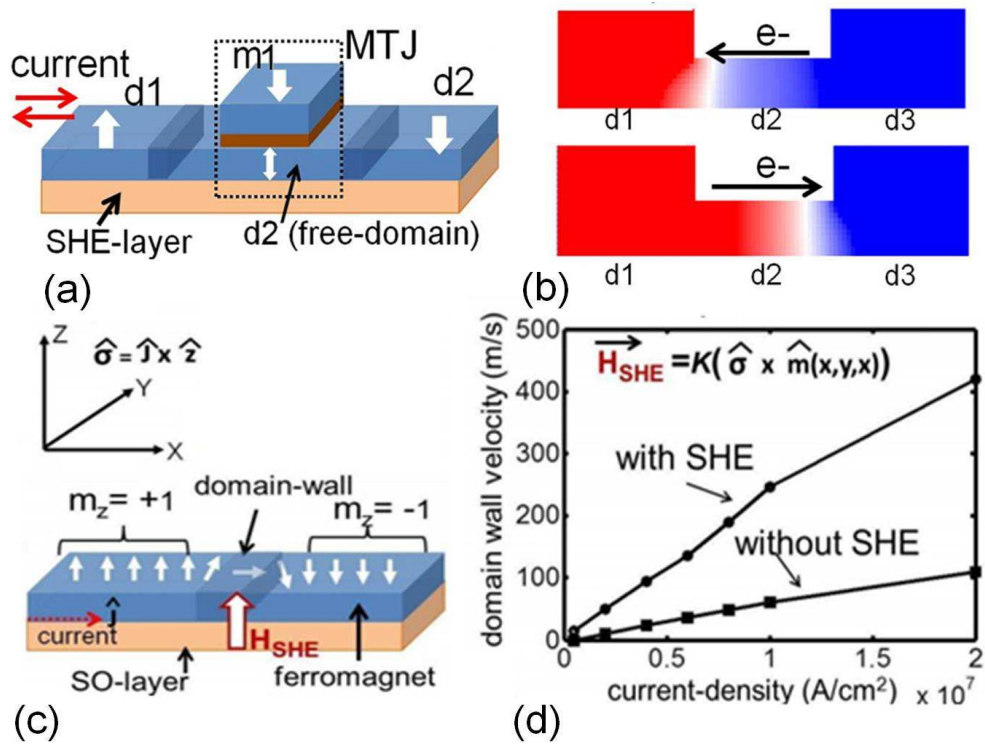

Fig. 3 (a) Three-terminal spin neuron based on domain wall magnet , (b) transient micro-magnetic simulation plots, (c) application of SHE assist for higher domain-wall (DW) speed, (d) DW-speed vs. current with and without SHE assist.

Fig. 3a shows a three terminal spin neuron based on domain wall magnet [14, 15]. It has a free magnetic domain $d_2$ which forms an MTJ with a fixed magnet $m_1$ at its top. The spin-polarity of $d_2$ can be written parallel or anti-parallel to the two fixed spin-domain $d_1$ and $d_3$, depending upon the direction of current flow between $d_1$ and $d_3$. Thus, this device can detect the direction or polarity (positive if going in and negative if going out of its input domain $d_1$) of current flow across its free domain. Hence this device can be used for current-mode thresholding operation [10]. The minimum magnitude of current flow required to flip the state of the free domain $d_2$ depends upon the critical current-density for magnetic domain wall motion across the free-magnetic domain $d_2$. Notably, domain-wall (DW) velocities of ~100m/s can be reached in magnetic nano-strips with current-density of ~$10^7$ A/cm$^2$ [14, 16, 17, 18]. Thus a spin neuron with 60nm long fee-layer with cross-section area of 20x2 nm$^2$ may be switched with a current of less than 10μA within 1ns [14, 22].

Recently, application of spin-orbital (SO) coupling in the form of Spin Hall Effect (SHE) has been proposed for low-current, high-speed domain-wall motion [19, 20, 21]. For Neel-type DW, SHE induced from an adjacent metal layer results in an effective magnetic-field ($H_{SHE}$) [19], that can be expressed as, $H_{SHE} = K(\sigma \times m)$. Here, $m$ denotes the magnetization of magnetic domains. $\sigma$ is a current-dependent vector defined as $\sigma = j \times z$, where, $j$ is the current vector (which can be positive or negative depending upon direction of current flow) and $z$ is the direction perpendicular to the magnetization plane (along easy axis). As shown in fig. 3a, $\sigma$ can be in-plan or out of plane of the figure, depending upon the direction of the current-flow. $K$ is a quantity dependent upon material parameters of the magnet and is proportional to the effective Spin-Hall angle, $\theta_H$ [19]. Notably, $\theta_H$ determines the effectiveness of the Spin-Hall interaction, larger $\theta_H$ implies larger effective torque due to SHE.

For a Neil-type domain wall shown in fig. 3a, the magnetization in the region of the domain wall lies along the length of the magnetic nano-strip [19]. For this configuration, the effective $H_{SHE}$ acting on the domain wall region can be visualized to be perpendicular to the plane of the magnet. The $H_{SHE}$ assists the non-adiabatic spin-torque (which results from the current-flow) acting on the domain-wall region. For a $\theta_H$ of 0.2, micromagnetic simulations showed an increase of ~5x in the domain-wall velocity for a given current density, due to the $H_{SHE}$ term (fig. 3d). This effect can be used to achieve higher switching speed for a given current, or, to reduce the required switching-current for a given switching-time for the free-domain in the spin-neuron [23].

In this work switching current threshold of ~2µA for 1 ns switching-speed has been chosen for a neuron with SHE-assisted free-domain size of $20 \times 2 \times 60 nm^3$, which corresponds to the current-density of $4MA/cm^2$. This dimension of the free-domain would offer an effective resistance of ~60Ω. The state of the free-domain can be sensed by injecting a small current across the high-resistance magnetic tunnel junction (MTJ) formed between $d_2$ and a fixed-magnet $m_1$.

Fig. 4 shows a spin-neuron with inputs synapses connected to domain-1. Domain-3 is connected to the ground potential. Due to low-resistance of the magneto-metallic write-path of the neuron, in absence of any input signal, the input terminal of the neuron is also clamped to the ground potential. This naturally fulfills the requirement of low impedance input-node along with a fixed input potential for the neuron device. Assuming a neuron with ~25 inputs and a maximum current of ~3µA per input, $|\Sigma(G_i V_i)_p| + |\Sigma(G_i V_i)_n|$ and $\Sigma(G_i V_i)$ come close to ~40µA and ~3µA. This implies an overall current-flow of ~3µA in and out of the DWM neuron (with resistance ~60Ω), which would result in a fluctuation of ~0.2mV at the input node. Thus even for input voltages as small as 10mV, the percentage fluctuation in the input-node-voltage can be less than ~2%. Moreover, it should be noted that, this fluctuation (positive or negative) is caused by the input itself. The net input-current injected into the neuron changes the input voltage in the direction determined by the larger of positive and negative current components (ie., according to the direction of the current flow at the neuron input). Hence it may not affect the final outcome unless it is large enough to reduce the current (difference between the positive ($\Sigma(G_i V_i)_p$) and the negative ($\Sigma(G_i V_i)_n$) components) injected into the neuron below its switching threshold (in this case designed to be ~2μA).

The state of the neuron's free layer (domain-2) can be detected using a high-resistance voltage-divider formed between a reference MTJ and the neuron MTJ, with the help of a simple CMOS inverter (fig. 4a). Thus the spin-neuron simultaneously provides transimpedance conversion for the input-current, thereby realizing the complete neuron equation in a single device.

The energy dissipation for the spin neuron has two components. First, the switching energy due to the static current flow between the input voltages and the neuron. These components equal to the product of the total input-current flowing across the synapses, the input-voltage levels and the neuron switching time. For an average of ~40μA of current flow across input voltage levels of $\pm$10mV for 1ns switching time, this component evaluates to ~0.4fJ. The noise considerations in the state of the art on-chip supply distribution schemes may limit the minimum input voltage levels that can be used. Even for $\pm$100mV of input levels, which might be more easily achievable, the first energy component is limited to ~4fJ, which is more than two orders of magnitude less than that obtained for the CMOS neuron. The second component of energy-dissipation in the spin-neuron can be ascribed to the MTJ-based read operation. A read current of ~0.3µA (~10% of neuron switching threshold) was found to be sufficient for 1ns read-speed. For a sensing supply voltage of 0.4V this would evaluate to ~0.12fJ. An additional ~0.2fJ of energy dissipation comes from the inverter's operation. Thus the total energy-dissipation in a spin-neuron for 1ns switching speed can be less than 1fJ. This leads to the possibility of three to four order of magnitude improvement in energy-delay product as compared to a conventional CMOS implementation. Apart from ultra-high energy efficiency, another attractive feature of the spin-neurons is their compactness. In the CMOS layer a compact CMOS inverter replaces an area consuming OPAMP. Hence, spin neurons can facilitate higher integration density for neural-network circuits.

A 3x3 neural-network circuit using spin neurons is shown in fig. 4b. The network has two conductances (that can be implemented using multi-level spintronic memristors) $G_{i+}$ and $G_{i-}$ for each input $in_i$. When an input is high (logic '1'), a voltage signal $+\Delta V$ and $-\Delta V$ are applied to the conductances $G_{i+}$ and $G_{i-}$ respectively, resulting in proportional current flow into the input terminal of the neuron, as shown in fig. 4b. The net current due to the $i_{th}$ input $in_i$, injected into the $j_{th}$ neuron, therefore, can be written as $\Delta V(G_{ii+}-G_{ii-})$. Thus, the input weights needed for the neurons can be obtained by programming $G_{i+}$ and $G_{i-}$ to appropriate states.

The write path of the neuron is connected to ground. Using Kirchhoff's law it can be visualized that the net current flowing into the input node of the neurons is given by the following equation:

$I_{sum} = \Sigma \Delta V((in_i(G_{ij+} - G_{ij-})))$ (3)

This expression is essentially same as the term within the braces in eq.1. The sign function over the current-mode summation is carried out by the spin-neurons, thus realizing the energy-efficient neural-network functionality. At the level of network-design, another noticeable advantage of spin-neurons is ultra-low energy-dissipation in cross-bar interconnects in the synapse network shown in fig. 4b. This results from the ultra low-voltage operation of entire network, facilitated by the spin neurons.

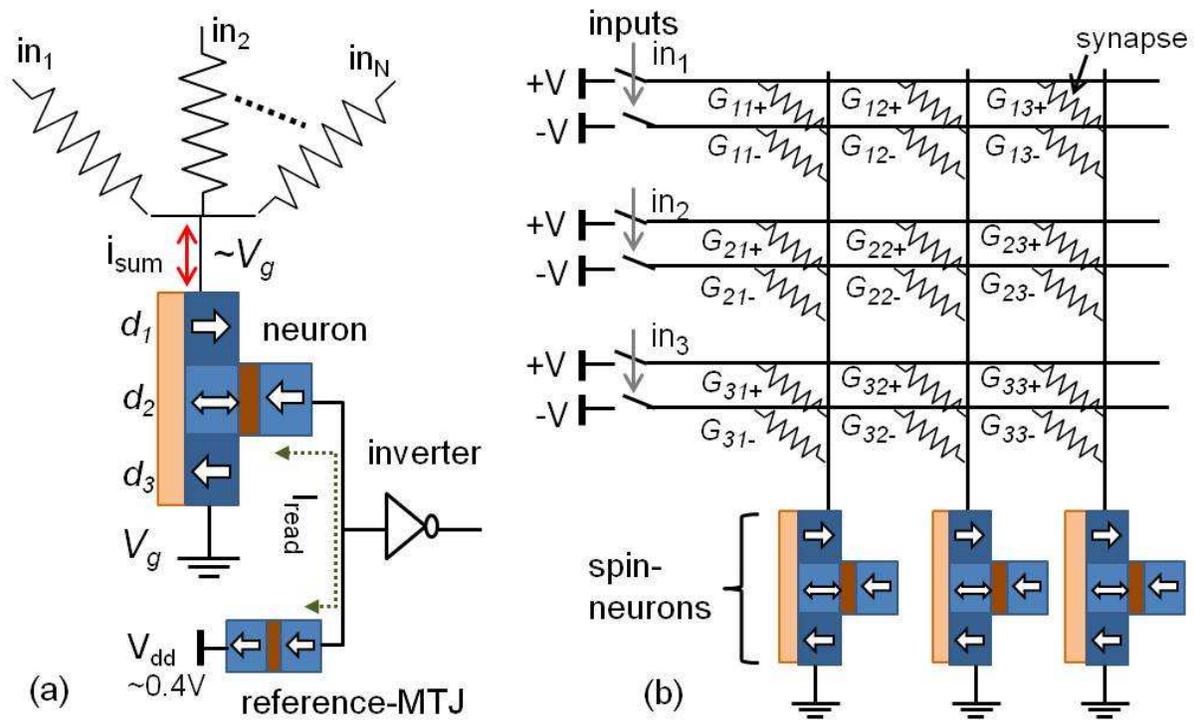

Fig. 4 (a) Spin-neuron connected with input synapses, (b) nerual-network circuit using spin-neurons

## IV. Conclusion

In this article we explained the rationale of using nano-scale spin-torque switches as "neurons" for the desgin of energy-efficient neuromorphic computers. Using simple device-circuit analysis we showed that spin neurons provide essential terminal characteritics like low input impedance and transfer chacteristics like high-transimpedance-gain and fast state-switching. These properties combined with ultra-low voltage operation of comapct spin-neurons can facilitate the design of ultra low-energy and high-performance bio-inspired computing systems.

**Acknowledgement:** This research was funded in part by NSF, SRC, DARPA, MARCO, and StarNet.